\documentstyle[preprint,aps,epsf]{revtex}
\begin{document}

\title{$K^+$ production in baryon-baryon and heavy-ion collisions}

\author{G. Q. Li$^{1}$, C. M. Ko$^{2}$, and W. S. Chung$^{2}$}

\address{$^1$Department of Physics, State University of 
New York at Stony Brook,\\
Stony Brook, New York 11794\\
$^2$Department of Physics and Cyclotron Institute,
Texas A\&M University, \\
College Station, Texas 77843}

\maketitle

\begin{abstract}

Kaon production cross sections in nucleon-nucleon, nucleon-delta 
and delta-delta interactions are studied in a boson exchange model. 
For the latter two interactions, the exchanged pion can be on-mass shell,
only contributions due to a virtual pion are included via the Peierls method 
by taking into account the finite delta width. With these cross 
sections and also those for pion-baryon interactions, 
subthreshold kaon production from heavy ion collisions 
is studied in the relativistic transport model. 

\end{abstract}

Kaon production in heavy-ion collisions at subthreshold 
energies continues to attract great interest, as it is 
expected to carry useful information not only on the 
initial collision dynamics but also on the nuclear equation 
of state at high density \cite{aich85,cass90,maru94,hart94,liko95a} 
and the kaon properties in dense matter \cite{fang94,koli96,kkl97}. 
However, to extract these information from the experimental 
data requires the use of transport models, in which elementary 
kaon production cross sections in hadron-hadron interactions are needed. 
At SIS energies ($\sim 1$ GeV/nucleon) the colliding system 
consists mainly of nucleons, delta resonances, and pions, so 
one needs kaon production cross sections in interactions between 
these particles. For kaon production in the nucleon-nucleon 
($NN$) interaction, there are only experimental data at high energies.
Thus various parameterizations have been introduced to make 
predictions at low energies. Furthermore, for processes involving 
baryon resonances, such as $N\Delta \rightarrow NYK$ and 
$N\Delta\to NYK$, which are unique in heavy-ion collisions 
and play important roles in subthreshold kaon production 
\cite{aich85,cass90,maru94,hart94,liko95a,fang94}, no experimental 
information is available. The usual prescription used in transport 
models is to assume that these cross sections are either the same as 
those in $NN$ interactions at same center-of-mass energies or somewhat
smaller using the scaling ansatz of Randrup and Ko based on isospin 
considerations \cite{rk80}.

There are attempts to evaluate the kaon production cross sections
using the one-boson exchange model 
\cite{wuko,laget,liko95,tsu97}, which has also been used 
frequently in calculating the pion \cite{mosel}, eta \cite{peters}, 
phi \cite{chung}, and dilepton \cite{schafer,haglin} production 
cross sections in $NN$ interactions. The extension of 
this model to particle production in nucleon-delta ($N\Delta$) and 
delta-delta ($\Delta\Delta$) interactions are straightforward, 
except for the complication that the exchanged pion can be 
on-mass shell, which then leads to a singular cross section.
Since the two-step process, which involves the decay of a delta 
into a nucleon and a pion and the subsequent interaction of the pion 
with another nucleon to produce a kaon, has already been included in the 
transport model, one should not include in $N\Delta$ and 
$\Delta\Delta$ interactions the contribution due to an on-shell 
pion. In Ref. \cite{liko95}, a complex pion self-energy based 
on a schematic delta-hole model has been introduced to evaluate 
the off-shell contribution. In a similar study of eta meson 
production from the $N\Delta$ interaction \cite{peters}, the 
imaginary part of the pion self-energy is determined from 
the pion-nucleon scattering cross section. Although 
the resulting cross sections are finite at finite densities,
they remain diverge as the density approaches zero. 

In heavy ion collisions, the delta-induced reaction is part of a 
more complex process that involves first the production of a delta in an $NN$ 
interaction, which can be either on or off its energy 
shell, and then its interaction with another nucleon.  
For kaon production, it was shown in Ref. \cite{bp96} that the 
contribution from an off-shell delta is much smaller than that 
from an on-shell one. Since an on-shell delta has a 
finite lifetime, its mass is thus complex with the imaginary 
part given by its width. Then energy-momentum conservation leads 
to a complex four momentum for the exchanged pion, which  
moves the singularity to the complex plane and thus makes 
the contribution from the exchange of an off-shell pion finite.   
Such a method was first introduced by Peierls \cite{Peierls} in  
understanding higher resonances production from the pion-nucleon reaction 
by first producing a delta and then followed 
by $\pi\Delta\to\pi N^*$ via a t-channel nucleon exchange. 
The singular amplitude at the nucleon pole was regularized 
by including the complex delta mass. In Ref. \cite{chung}, 
this method was used to evaluate the
phi meson production cross section in $N\Delta$ and
$\Delta\Delta$ interactions. 
This physical region singularity in the scattering of particles with
finite lifetimes appears also in muon collider physics \cite{muon}, 
where a singularity appears in the process $\mu^{+} \mu^{-} 
\rightarrow e \bar \nu W^{+}$, 
as the exchanged neutrino can be on-shell due to the decay 
$\mu\to e \bar \nu \nu_{\mu}$. In Ref. \cite{muon}, this singularity 
has also been regularized via the Peierls method by including 
the width of the muon.

The elementary kaon production cross sections in baryon-baryon
interactions have been studied in Ref. \cite{liko95} based
on the pion and kaon exchange model. By adjusting two cut-off
parameters $\Lambda _\pi$ and $\Lambda _K$ in the form factors at
the $\pi NN$ and $KN\Lambda$ vertices, respectively, very good 
agreements with experimental kaon production
cross sections in the $NN$ interaction have been obtained.
In Fig. 1, results for $pp\rightarrow p\Lambda K^+$ are 
compared with the experimental data and various parameterizations.
Open circles are old data from the compilation by
Baldini {\it et al,} \cite{bald88}, while the solid circle
is the data from the recent experiment at COSY for proton-proton 
collisions at 2 MeV above the kaon production threshold \cite{bale96}. 
It is seen that our model provides a good description of both 
old and new data. The parameterization of Randrup and Ko \cite{rk80} 
is shown in the figure by the dashed curve. It describes
reasonably well the experimental data at high energies
(open circles), which were available when it was introduced.
However, it significantly overestimates the COSY data
as a result of its incorrect threshold behavior. 
The parameterization of Sch\"urmann and Zwermann \cite{sch87} 
is shown in the figure by the dash-dotted curve. Its assumption 
of a quartic dependence on $p_{\rm max}$ (the maximum momentum
of kaon at a given center-of-mass energy) does not agree with 
the experimental data near the threshold either. The dotted 
curve in the figure comes from Cassing 
{\it et al.} \cite{cass97}, which is also based on
the boson-exchange model.

In evaluating the kaon production cross sections in $N\Delta$ and 
$\Delta\Delta$ interactions by the Peierls method, the pion 
propagator is modified by the imaginary delta energy due to 
its width $\Gamma_\Delta$,
i.e., 
\begin{equation}\label{pion}
   \frac{1}{\left(p_{\Delta}-p_{N} \right)^{2} - m_{\pi}^{2} }
   \longrightarrow
   \frac{1}{\left(p_{\Delta}-p_{N} \right)^{2} - m_{\pi}^{2}
              -i m_{\Delta} \frac{(E_{\Delta}-E_{N})}{E_{\Delta}} 
               \Gamma_{\Delta} },
\end{equation}
where $p_N$ and $p_\Delta$ are four momenta of the 
final-state nucleon and initial-state delta resonance, 
respectively; $E_N$ and $E_\Delta$ are their energies in the 
center-of-mass frame. The off-shell pion contribution is then 
obtained by including only the real part of Eq. (\ref{pion}).
In nuclear medium, the delta width is expected to change,
and also the exchanged pion can acquire additional imaginary 
self energy due to interactions. Such medium-dependent effects 
are interesting but require further study and are thus
neglected here.

Results for the isospin-averaged kaon production cross sections
for both $N\Sigma K$ and $N\Lambda K$  
final states are shown in Fig. 2, together with those from the 
parameterization of Randrup and Ko \cite{rk80}.
Near the threshold, because of their incorrect threshold
behavior (see Fig. 1), the Randrup-Ko cross sections
are larger than our microscopic results for all the channels.
In the energy region most relevant for kaon production
at SIS energies, our results for $N\Delta$ and $\Delta\Delta$
channels are larger than those of Randrup and Ko.

To apply these cross sections in the transport model, we have 
parameterized our theoretical results in terms of 
$\sqrt s-\sqrt {s_0}$, with $\sqrt s$ and $\sqrt {s_0}$ denoting,
respectively, the available energy and the threshold,
\begin{eqnarray}\label{fit}
\sigma _{BB\rightarrow NYK} = {a(\sqrt s-\sqrt {s_0})^2\over
b+(\sqrt s-\sqrt {s_0})^x} ~{\rm mb}.
\end{eqnarray}
The fitted parameters $a$, $b$ and $x$ for six channels are
listed in Table 1. In addition, we also include kaon production 
from pion-baryon interactions with cross sections taken from 
Ref. \cite{fae94}, which were also parameterized in terms of 
$\sqrt s-\sqrt {s_0}$. Furthermore, we include processes
with one or two pions in the final states. The cross sections
for these processes are obtained from parameterizations
based on experimental data \cite{gqli}.
When medium effects on kaons are considered, the threshold is
calculated with in-medium masses. In addition, kaons
also propagate in their mean field potentials \cite{likoli}.

Kaon properties in dense matter have been extensively studied;
for a recent review see Ref. \cite{lee96}.
These studies indicate that the kaon 
energy in medium can be expressed as 
\begin{eqnarray}
\omega _{K}=\left[m_K^2+{\bf k}^2-a_{\bar K}\rho_S
+(b_K \rho_N )^2\right]^{1/2} + b_K \rho_N,
\end{eqnarray}
where $b_K=3/(8f_\pi^2)\approx 0.333$ GeV fm$^3$ represents
the vector repulsion, while $a_K$ determines the strength 
of the attractive scalar potential. If one considers only the
Kaplan-Nelson term, then $a_K=\Sigma _{KN}/f_\pi ^2$,
with $\Sigma _{KN}$ being the $KN$ sigma term. In the same order, 
there is also an energy-dependent range term which cuts down the 
scalar attraction \cite{lee96}. Since both $\Sigma_{KN}$ and 
the range term are not very well determined, we treat $a_K$ 
as a free parameter. Taking $a_K\approx 0.22~ {\rm GeV}^2~{\rm fm}^3$, 
then a kaon feels a repulsive potential of about 20 MeV at 
normal nuclear matter density, in rough agreement
with the prediction of impulse approximation using the
$KN$ scattering length in free space \cite{likoli}. 

Extensive transport model calculations for subthreshold
kaon production have been carried out recently in
\cite{cass97,gqli,cass97a}. Here we shall concentrate on
$K^+/\pi^+$ ratio in Au+Au collisions at 1 AGeV, as measured
by KaoS Collaboration \cite{kaos94,kaos96}. We first show
in Fig. 3 the $\pi^+$ transverse momentum spectra in Au+Au
collisions at 1 AGeV. The results from our transport 
model (histogram) are in very good agreement with the recent 
data from both the FOPI Collaboration \cite{pelte97} (open cicles)
and the Kaos Collaboration \cite{kaos96} (open squares).

In the upper panel of Fig. 4 we show the $K^+$ yield, 
normalized by the number of participant nucleons $A_{\rm part}$, 
as a function of $A_{\rm part}$. In both cases with and 
without kaon medium effects, the ratio increases more than 
a factor of two from peripheral to central collisions. 
On the other hand, the ratio $\pi^+/A_{\rm part}$ stays 
almost a constant of 0.03 in the entire impact parameter 
range. The stronger dependence of the $K^+$ yield than the 
pion yield on the centrality of the collision
indicates that secondary processes involving pions and
delta resonances are extremely important for subthreshold particle
production. Indeed, for central Au+Au collisions, the kaon yield
from pion-baryon collisions accounts for more than 60\% of the
total kaon yield, as was also found in Ref. \cite{cass97a}.
Furthermore, the $N\Delta$ channel is found to be the most
important one among all the baryon-baryon contributions.

In the lower window of Fig. 4 we show the
ratio $K^+/\pi^+$ as a function of $A_{\rm part}$. It is
seen that our results with kaon medium medium effects
are in agreements with the latest KaoS data shown
by solid squares \cite{kaos96}. The results without kaon medium effects
overestimate this ratio. Also shown in the figure by open squares
are the previous KaoS data published in Ref. \cite{kaos94}, which
seem to be better described by the scenario without the
kaon medium effects (which implies a complete cancellation
of the scalar and vector potentials in Eq. (3)). 

In summary, extending our previous work on phi meson
production cross sections in baryon-baryon collisions \cite{chung},
we have evaluated the kaon production cross sections 
in $N\Delta$ and $\Delta\Delta$ interactions using the 
Peierls method by including the finite delta width. 
Using them in the relativistic transport model, we have 
studied kaon production, in particular the $K^+/\pi^+$ ratio,
in Au+Au collisions at 1 AGeV.
We found that the latest KaoS data on this ratio are
consistent with a weak repulsive kaon potential.

\vskip 1cm
We thank Gerry Brown, C.-H. Lee and P. Senger for useful discussions.
G.Q.L. was supported in part by the Department
of Energy under Contract No. DE-FG02-88ER40388,
while C.M.K. and W.S.C. were supported in part by the National 
Science Foundation under Grant No. PHY-9509266.

\newpage

{\bf Table 1} Fitted parameters in Eq. (2).

\vskip 0.5cm

\begin{center}
\begin{tabular}{ccccccc}
\hline
& $NN\rightarrow N\Lambda K$ & $NN\rightarrow N\Sigma K$ &
$N\Delta\rightarrow N\Lambda K$ & $N\Delta\rightarrow N\Sigma K$ &
$\Delta\Delta\rightarrow N\Lambda K$ & $\Delta\Delta\rightarrow N\Sigma K$\\
$a$ & 0.0865 & 0.1499 & 0.1397 & 0.3221 & 0.0361 & 0.0965 \\
$b$ & 0.0345 & 0.167  & 0.0152 & 0.107  & 0.0137 & 0.014 \\
$x$ & 2.0    & 2.4    & 2.3    & 2.3    & 2.9    & 2.3   \\
\hline
\end{tabular}
\end{center}

\newpage  

\centerline{\bf Figure Captions}

Fig. 1: Kaon production cross section for $pp\rightarrow p\Lambda K^+$.
The solid curve is the results from the boson-exchange model of Ref.
\cite{liko95}. The dashed, dash-dotted, and 
dotted curves are, respectively, the parameterizations of Randrup and Ko
\cite{rk80}, Sch\"urmann and Zwermann \cite{sch87},
and Cassing {\it et al.} \cite{cass97}. Open circles
are experimental data from Baldini {\it et al.} 
\cite{bald88}, while the solid circle is from
Ref. \cite{bale96}.

Fig. 2: Isospin-averaged kaon production cross sections in
$NN$, $N\Delta$, and $\Delta\Delta$ interactions. Solid curves
are from this work, and dotted curves are from the 
Randrup-Ko parameterization \cite{rk80}.

Fig. 3: $\pi^+$ transverse momentum spectra in Au+Au collisions
at 1.0 AGeV. The histogram gives our theoretical results, while
the circles and squares are the experimental data from
the FOPI \cite{pelte97} and KaoS \cite{kaos96}
collaborations, respectively.

Fig. 4: Upper window: $\pi^+$ and $K^+$ yields normalized 
by the participant nucleon number $A_{\rm part}$ in 
Au+Au collisions at 1 AGeV. Lower window: $K^+/\pi^+$ ratio 
as a function of the participant nucleon number.

\newpage

\begin{figure}
\begin{center}
\vfill
\mbox{\epsfxsize=12truecm\epsffile{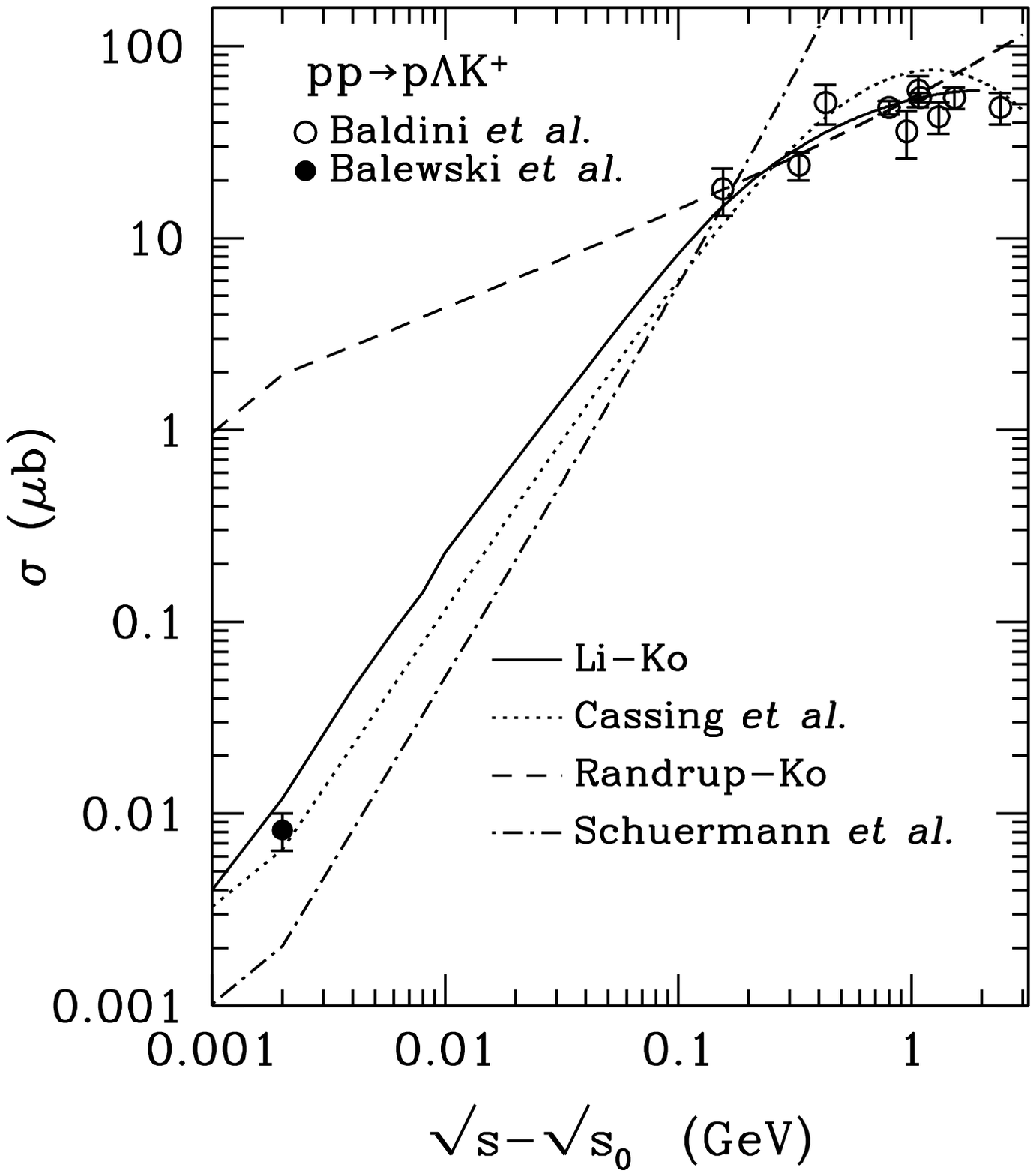}}
\vfill
\end{center}
\end{figure}

\newpage
\begin{figure}
\begin{center}
\vfill
\mbox{\epsfxsize=12truecm\epsffile{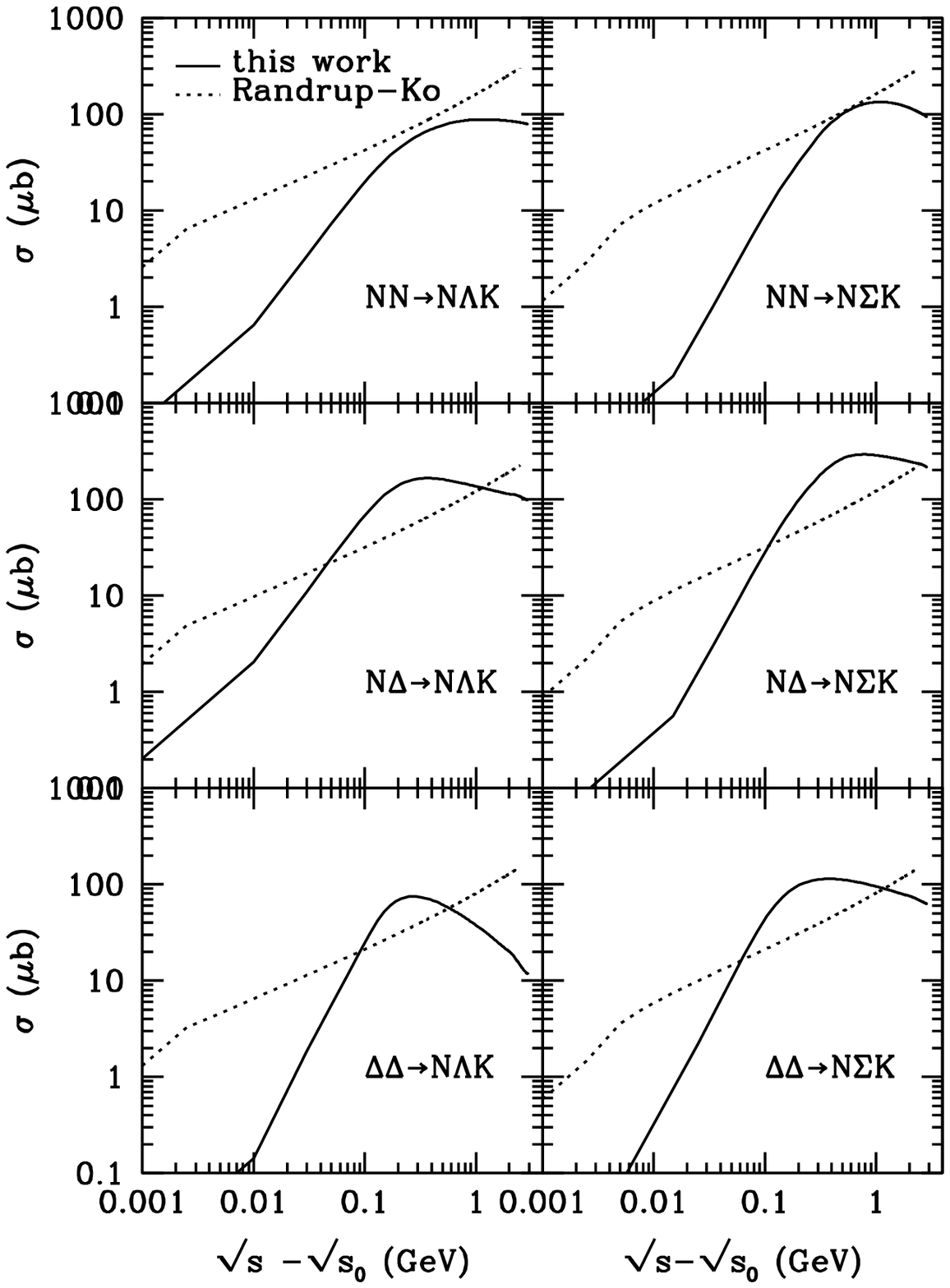}}
\vfill
\end{center}
\end{figure}

\newpage
\begin{figure}
\begin{center}
\vfill
\mbox{\epsfxsize=12truecm\epsffile{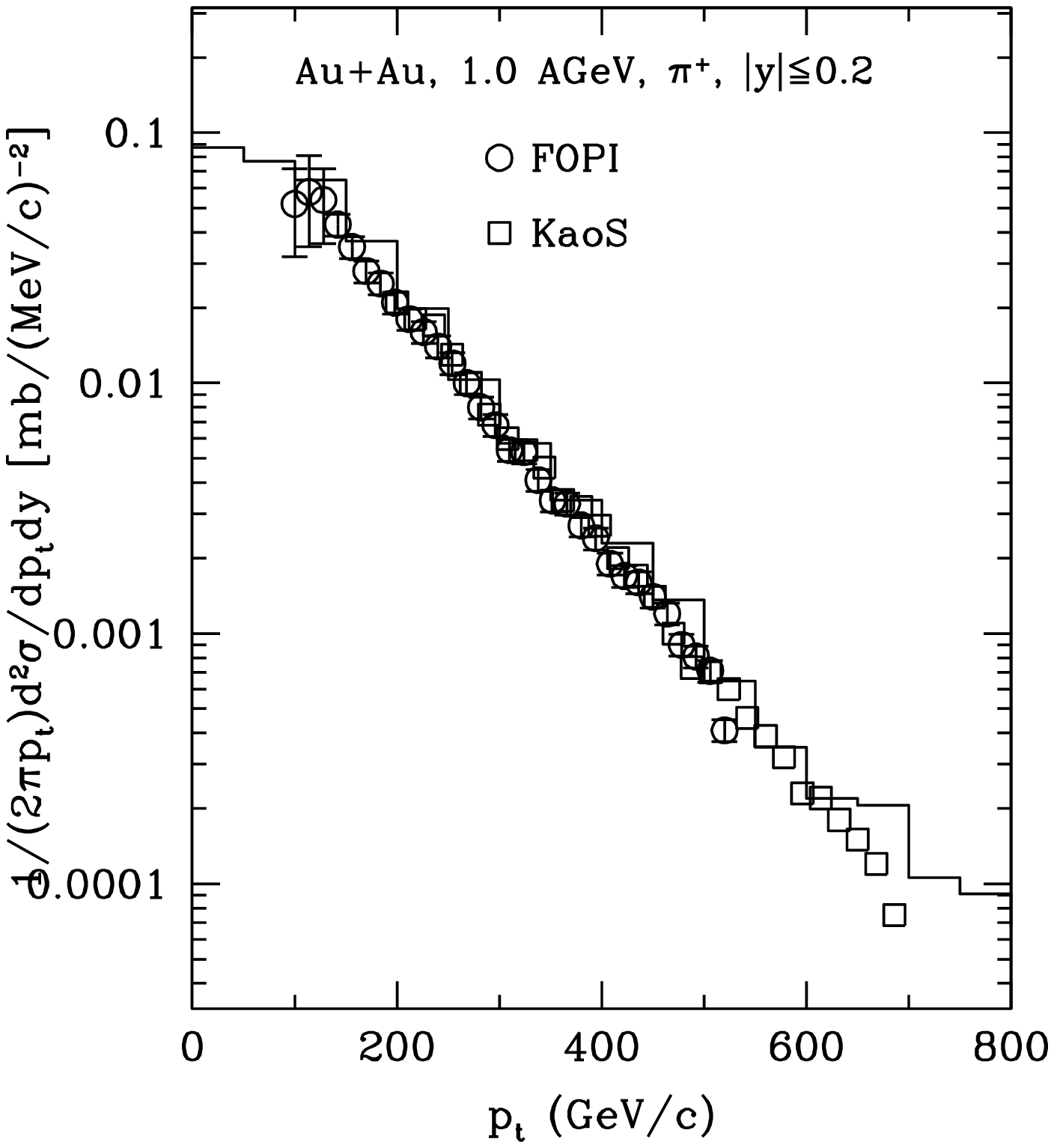}}
\vfill
\end{center}
\end{figure}

\newpage
\begin{figure}
\begin{center}
\vfill
\mbox{\epsfxsize=12truecm\epsffile{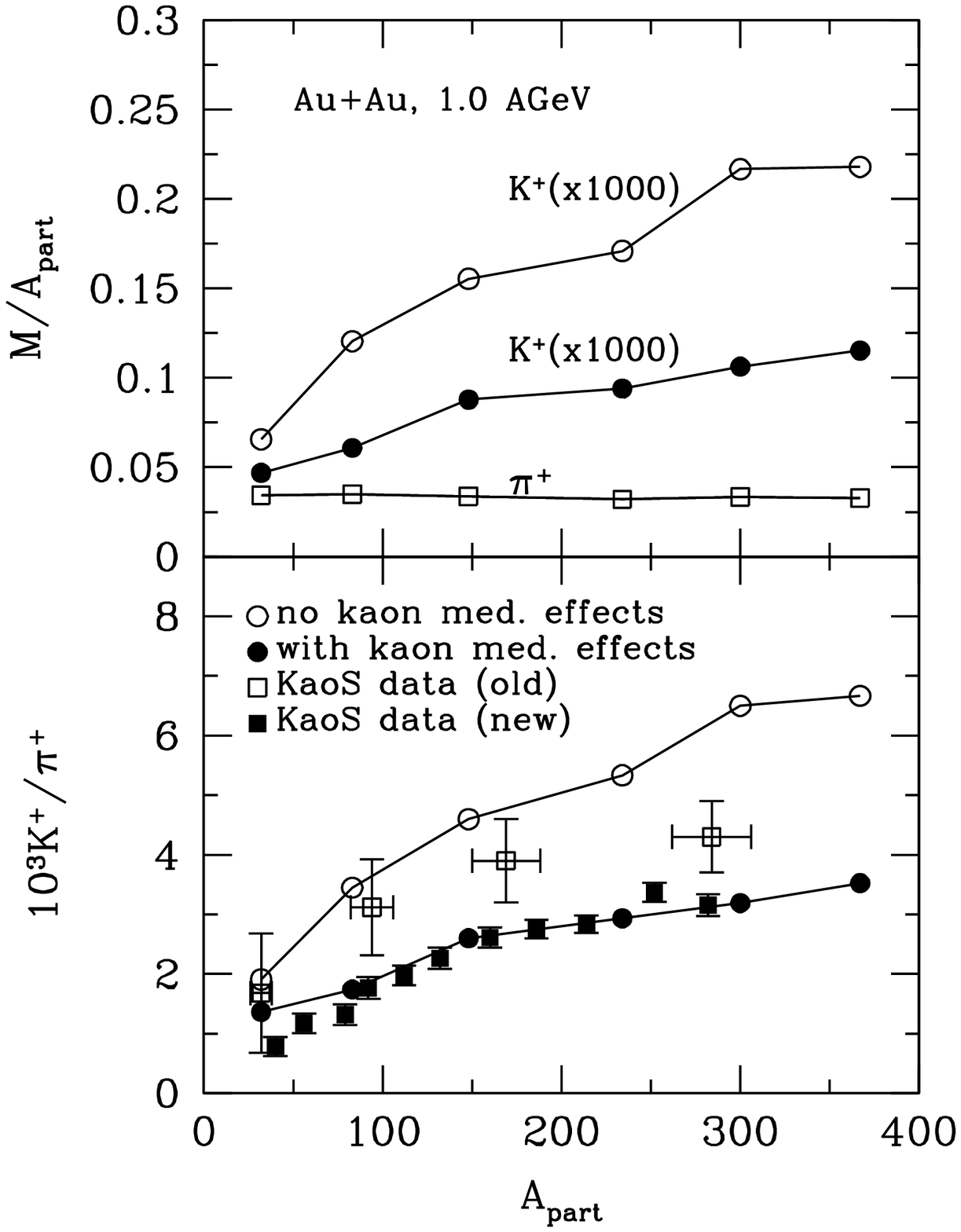}}
\vfill
\end{center}
\end{figure}

\end{document}